\documentclass[sigconf,natbib=false]{acmart}

\AtBeginDocument{%
 }

\setcopyright{acmcopyright}
\copyrightyear{2022}
\acmYear{2022}
\acmConference[MADiMa '22]{Proceedings of the 7th International Workshop on
  Multimedia Assisted Dietary Management}{October 10, 2022}{Lisboa, Portugal.}
\acmBooktitle{Proceedings of the 7th International Workshop on Multimedia
  Assisted Dietary Management (MADiMa '22), October 10, 2022, Lisboa, Portugal}
\acmPrice{15.00}
\acmDOI{10.1145/3552484.3555746}
\acmISBN{978-1-4503-9502-1/22/10}

\RequirePackage[
  datamodel=acmdatamodel,
  style=acmnumeric
]{biblatex}

\usepackage{graphicx}
\graphicspath{{figures/}}

\usepackage{xcolor}

\usepackage{hyperref}
\usepackage{booktabs}

\begin{document}

\title{Chewing Detection from Commercial Smart-glasses}

\author{Vasileios Papapanagiotou}
\email{vassilis@mug.ee.auth.gr}
\orcid{0000-0001-6834-5548}
\affiliation{%
  \institution{Multimedia Understanding Group\\Electrical and Computer Engineering Dpt.\\Aristotle University of Thessaloniki}
  \city{Thessaloniki}
  \country{Greece}
  \postcode{54124}
}

\author{Anastasia Liapi}
\email{liapianast@ece.auth.gr}
\affiliation{%
  \institution{Electrical and Computer Engineering Dpt.\\Aristotle University of Thessaloniki}
  \city{Thessaloniki}
  \country{Greece}
  \postcode{54124}
}

\author{Anastasios Delopoulos}
\email{antelopo@ece.auth.gr}
\orcid{0000-0001-8220-8486}
\affiliation{%
  \institution{Multimedia Understanding Group\\Electrical and Computer Engineering Dpt.\\Aristotle University of Thessaloniki}
  \city{Thessaloniki}
  \country{Greece}
  \postcode{54124}
}

\renewcommand{\shortauthors}{Papapanagiotou et al.}

\begin{abstract}
  Automatic dietary monitoring has progressed significantly during the last
  years, offering a variety of solutions, both in terms of sensors and
  algorithms as well as in terms of what aspect or parameters of eating behavior
  are measured and monitored. Automatic detection of eating based on chewing
  sounds has been studied extensively, however, it requires a microphone to be
  mounted on the subject's head for capturing the relevant sounds. In this work,
  we evaluate the feasibility of using an off-the-shelf commercial device, the
  Razer Anzu smart-glasses, for automatic chewing detection. The smart-glasses
  are equipped with stereo speakers and microphones that communicate with
  smart-phones via Bluetooth. The microphone placement is not optimal for
  capturing chewing sounds, however, we find that it does not significantly
  affect the detection effectiveness. We apply an algorithm from literature with
  some adjustments on a challenging dataset that we have collected in
  house. Leave-one-subject-out experiments yield promising results, with an
  F1-score of $0.96$ for the best case of duration-based evaluation of eating
  time.
\end{abstract}



\maketitle

\section{Introduction}

Detection of chewing sounds is one of the first approaches that have been
studied in the field of automatic dietary monitoring
\cite{amft2005analysis}. The idea is to capture (by audio) the distinct sound
that occurs when food is crashed between the teeth. Different placements of the
microphone have been considered, but the one that seems naturally advantageous
is the outer in-ear canal, as it captures the chewing sounds naturally amplified
by the skull, while external sounds are attenuated. Typically, microphones have
been mounted on custom housing to support this placement, which creates the need
for custom-made hardware.

In recent years, various approaches have also emerged, focusing on different
sensor solutions such as detection of swallowing sounds \cite{amft2006methods,
  rayneau2021automatic, aboofazeli2009swallowing}. The naturally optimal
placement of the microphone for this task is close to the neck, where swallowing
sounds originate from. Other approaches attempt to leverage the power of modern
wearables, such as smart-watches, by using the inertial sensors
(i.e. accelerometer and optionally gyroscope) that are commonly found on such
devices, in order to detect and identify eating gestures
\cite{kyritsis2021assessment, kyritsis2021data, heydarian2020deep},

Other approaches rely on the availability of cameras to identify food type
\cite{anthimopoulos2014food, anthimopoulos2015computer} from single plate
photographs, or to directly segment food images into food components
\cite{aslan2018semantic, ciocca2019evaluating}. It is also possible to estimate
food volume using a depth camera \cite{lo2018food} and the caloric content based
on a photograph using a reference object \cite{ege2019association}. In
\cite{dehais2017two}, authors reconstruct a 3D model of the food that is placed
on a plate in order to estimate food volume; however, their approach requires
two different views (photographs) of the plate. While such approaches can
clearly extract very detailed information, they require the active participation
of the user by taking a photograph and triggering the analysis, or in some cases
specialized cameras (such as depth camera) or multiple views. However, combining
with monocular depth-estimation methods \cite{ming2021deep} can help reduce user
and hardware input requirements and simplify the estimation process.

Besides smart-watches, another wearable that has received attention for dietary
monitoring is the smart-glasses. Smart-glasses are mostly in experimental state
currently, however, many alternative sensors have been mounted and evaluated in
the literature. In \cite{rui2016diet}, authors use 3D-printed glasses that
incorporate an electromyography (EMG) sensor and evaluate its potential on
detecting chewing and identifying food types, on a dataset of eight participants
and five food types. The best reported chewing detection effectiveness is $0.8$
for both recall and precision, and classification accuracy is reported in the
range of $0.64$ to $0.84$. In \cite{rui2016bite}, they combine it with a
vibration sensor. In \cite{rui2018monitoring}, authors use 3D printed
smart-glasses with a bilateral EMG sensor to detect chewing. They evaluate both
in lab and free-living conditions. Results for free-living conditions in chewing
detection achieve $0.79$ recall and $0.77$ precision.

In \cite{mertes2018detection}, authors opt for a 3D accelerometer sensor (from
Shimmer), mounted with plastic straps on regular glasses, as a prototype
device. They propose an algorithm based on feature extraction and
classification, using either support vector machines (SVMs) or random forests,
targeting three classes: chewing, non-chewing, and walking (as a challenging
counterpart for chewing). They evaluate on a dataset from five volunteers, and
achieve a classification accuracy of $0.74$, however, the prior probability of
each class in the dataset is not given.

A pair of glasses with mounted EMG and additional electronics to support
communication over Bluetooth with a smart-phone is used in
\cite{huang2017your}. The proposed algorithm is made of two parts: one that runs
on the electronics mounted on the glasses and detects eating periods
(vs. non-eating periods) and one that runs on a more powerful processing unit
(after the data transfer via Bluetooth) that detects individual chews. Chewing
detection is evaluated on dataset of four individuals that eat, drink, and talk,
while sited for $40$ minutes each, and achieves $0.96$ accuracy.

The authors of \cite{chung2017glasses} propose a different approach by using 3D
printed glasses that incorporate load cells. The principle of operation is that
during chewing, the temples of the glasses (the part that is closer to the ear)
is slightly pressed outward, and this increases the load on the cell that is
placed at the hinge. The algorithm includes extraction of both time-domain and
frequency-domain features and training of an SVM classifier with an radial-basis
function (RBF) kernel. Authors evaluate on a dataset of $10$ subjects on $6$
classes that include left/right side chewing, left/right side winking, talking,
and head moving, and report an average F1-score of $0.94$.

In this current work, we try to combine the advantages of multiple approaches
into a single solution. We employ audio-based chewing detection, as it can
provide very accurate results, detailed (per chew) detection, can be used to
identify food type \cite{mirtchouk2016automated}, food texture
\cite{papapanagiotou2021recognition}, and even bite weight \cite{amft2009bite,
  papapanagiotou2021bite}, and also does not require active user input. We
employ a commercial, off-the-shelf device to remove the requirement for
specialized/custom-made hardware. We opt for a pair of Bluetooth-enabled
smart-glasses by Razer. Finally, we employ a chewing-detection algorithm from
literature that has been previously employed on a sensor that combines audio,
photoplethysmography, and acceleration signals \cite{papapanagiotou2017}; the
algorithm is resilient to the challenges that this device introduces, such as
the non-optimal placement of microphones. To evaluate our work, we conducted a
data-collection trial.

The rest of this work is organized as follows. Section \ref{sec:materials}
describes the sensor, data-collection process, and the algorithm for chewing
detection. Section \ref{sec:experimental} presents the evaluation framework and
results and discusses them. Finally, section \ref{sec:conclusions} concludes the
paper.

\section{Materials and methods}
\label{sec:materials}

\subsection{Hardware and data-collection}

\begin{figure}
  \centering
  \includegraphics[width=\columnwidth]{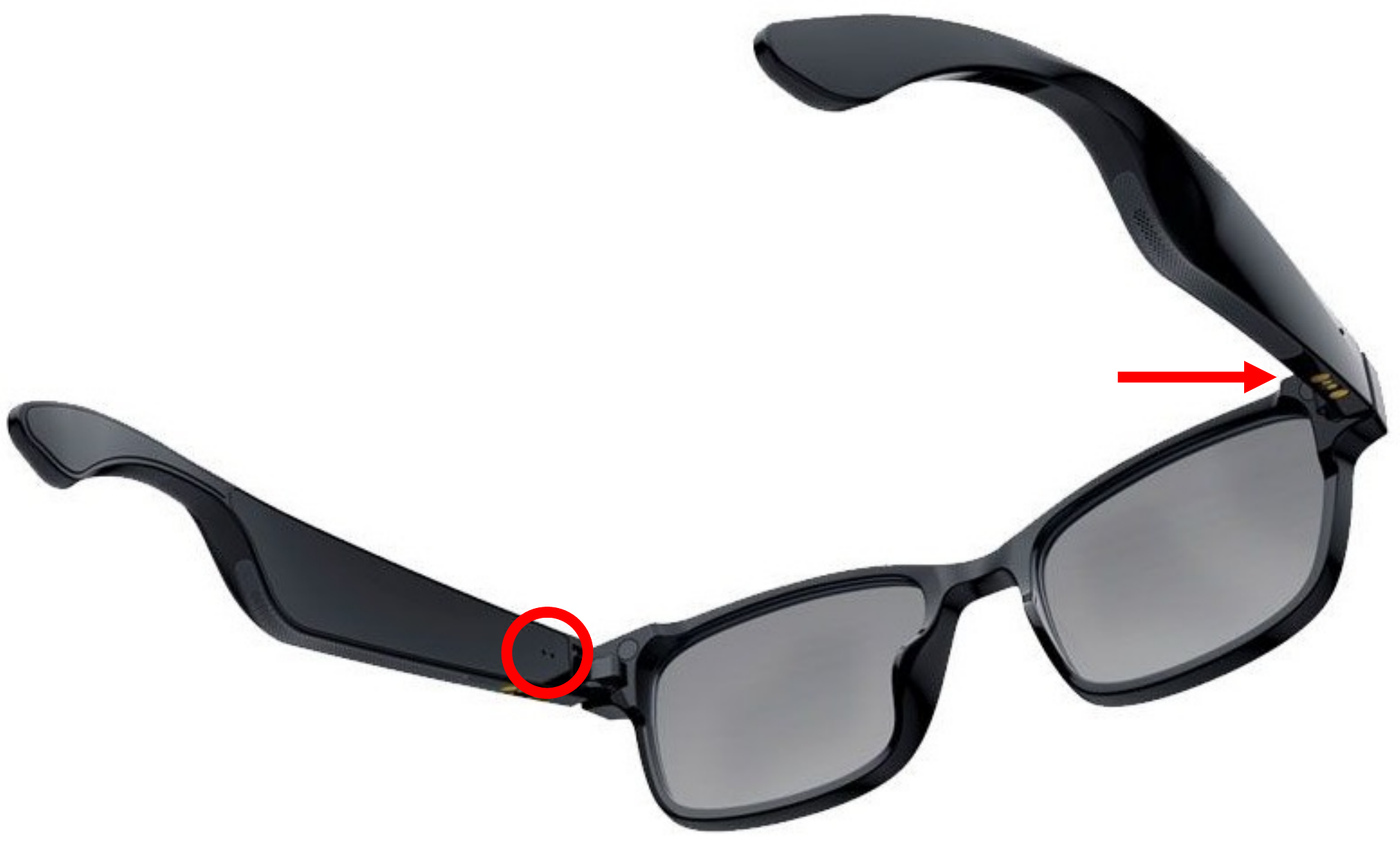}
  \caption{The commercial smart-glasses that are used for collecting audio,
    Razer Anzu. (Photo from
    \url{https://www.razer.com/mobile-wearables/razer-anzu-smart-glasses}). There
    are two microphones marked by the red circle (left side) and arrow (right
    side, microphone not visible due to perspective).}
  \label{fig:anzu}
\end{figure}

The hardware we use in this work for recording audio is the commercially
available off-the-shelf smart-glasses by Razer, the Razer Anzu (Figure
\ref{fig:anzu}). They are regular glasses (which can also be used as sunglasses)
that include stereo speakers and microphones, and communicate with smart-phones
and laptops wirelessly via Bluetooth. The microphones are placed on the glasses
temples, facing inside, and very close to the glasses end-pieces. They are
essentially facing the subject's eyes from the outer sides. Figure
\ref{fig:anzu} marks the microphone placement with the red symbols: the
microphone on the left side is at the two, very small, dark dots inside the red
circle; the red arrow points to the right microphone which is not visible due to
perspective.

The audio recorded from the smart-glasses is stereo (2 channels). We sample at
$16$ kHz and we use the Advanced Audio Coding (AAC) standard for
compression. The compression yields files with size less than $6$ MBs per hour
which permits uploading the audio files to a server for processing and
detection.

Based on existing literature \cite{amft2005analysis}, chewing sounds travel well
through the skull and are best captured with microphones placed inside the ear's
outer canal, facing inwards (this placement also naturally attenuates external
sounds). However, continuous use of in-ear microphones can create discomfort to
some users \cite{boer2018splendid}. On the other hand, the Razer Anzu
smart-glasses are worn as common glasses and are thus more comfortable. They are
also available to buy commercially, which eliminates the need for specialized
hardware. The only drawback to opting for the smart-glasses is that the
placement of microphones is not optimal for chewing detection. Indeed,
microphone placement in the Razer Anzu is oriented towards capturing speech, as
their target use-case is to be used as a headset. This creates a bigger
challenge for the chewing detection algorithm (compared to using in-ear
microphones).

\subsection{Dataset}

To record audio from the smart-glasses we have developed an application for
Android smart-phones that can connect to the smart-glasses via Bluetooth and
record audio in real-time. The audio files are stored on the phone and we
manually gather them to a computer for analysis and processing. However, it is
technically feasible to upload them automatically to a server, given their
relatively small size.

\begin{figure}
  \centering
  \includegraphics[width=.7\columnwidth]{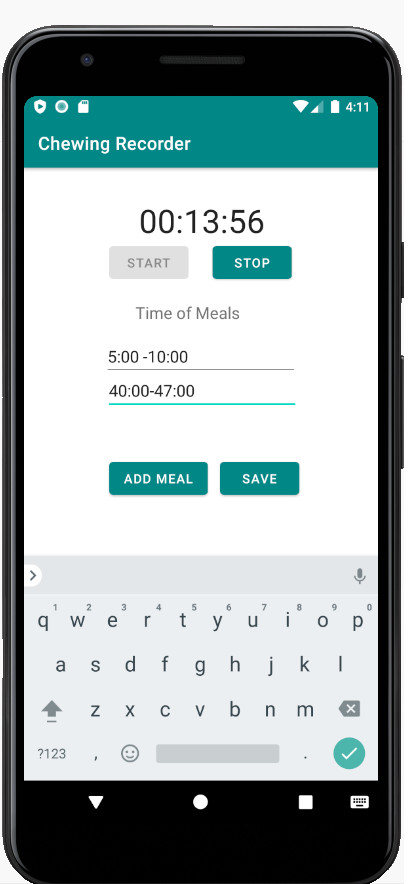}
  \caption{Main interface of the Android application that was developed for
    recording audio from the smart-glasses. It also enables the user to provide
    start and stop timestamps for eating sessions (meals).}
  \label{fig:app}
\end{figure}

The application includes an interface (Figure \ref{fig:app}) for the subject to
annotate the start and stop timestamps of eating sessions (meals). This helps us
to manually annotate individual chews, which is required for training.

To analyze and evaluate our method we collected a dataset of audio data.  A
total of $5$ subjects ($1$ male and $4$ female, age range $23$ to $28$)
participated in the collection process. Subjects had no reported or diagnosed
medical issues relevant to eating and digestion. Each recording lasted
approximately an hour, and we collected $6$ such recordings (one subject
contributed twice). Subjects were instructed to perform the following
activities: eating, talking, walking, and resting. Subjects were free to perform
these activities in any order, and in any way they wanted. The only instructions
that were given were to perform each activity for at least $10$ minutes, and
have at least $2$ eating sessions (so at least $20$ minutes of eating).

The recorded audio signals were inspected in
Audacity\footnote{\url{https://www.audacityteam.org/}}. We used the eating
session annotations that were provided by the subjects via the Android
application interface as guides; based on the session annotations we manually
annotated each individual chew with start and stop timestamps. In total, the
dataset contains $6$ hours and $14$ minutes of audio, $12$ eating sessions
(meals), and $9,562$ individual chews. For the chew duration, the mean is
$0.348$ sec and the standard deviation is $0.046$ sec. Consumed food types
include bread, cucumber, ice-cream, snack bars, and biscuits.

The study was approved by the Ethics committee of the Aristotle University of
Thessaloniki (151279/2022). The Ethics committee has also approved the public
publishing of derivatives of the dataset (such as mel-frequency cepstral
coefficients) along with the manual ground-truth annotations and
demographics. All participants were informed about the study, the use of their
data, and potential public publishing, and have signed a written consent form.

\subsection{Detection algorithm}
\label{sec:algorithm}

To detect chewing activity, we employ an algorithm from literature
\cite{papapanagiotou2017}. The algorithm includes several steps: a
pre-processing filter, feature extraction from short, overlapping windows,
training of a binary SVM classifier (chewing vs. non-chewing), and SVM-score
post-processing smoothing. The detector yields a sequence of binary labels which
correspond to chewing activity.

Based on that, individual chews, chewing bouts, and then eating sessions can be
obtained, using a set of heuristic rules. The code for aggregating the
classification labels to meals is available
online\footnote{\url{https://github.com/mug-auth/chewing-detection-challenge}}
from \cite{papapanagiotou2017splendid}.

The biggest challenge for detecting chewing sounds in this work is the different
placement of the microphones, compared to the ``traditional'' in-ear placement
commonly found in literature. The natural amplification of chewing sounds
through the skull is not present here, while talking sounds are better amplified
(since this is the target use-case of the smart-glasses). Based on that, we
choose the algorithm of \cite{papapanagiotou2017} as it is resilient to both
challenges.

First, each audio window is normalized before extracting the features by
dividing each audio sample with the standard deviation of the audio samples
within said window. This step, in combination with the high-pass filtering (with
a very low cut-off point) during the pre-processing stage of the algorithm,
essentially result in the standardization of each audio window (i.e. forces a
mean of $0$ and a standard deviation of $1$). It should be noted that based on
our observations, the captured audio is already zero-mean, so the application of
the high-pass filter is redundant.

Second, the algorithm uses a small set of carefully selected features, some of
which are exceptionally good at differentiating talking from other sounds, such
as the fractal dimension \cite{papapanagiotou2015fractal}, the condition number
of the auto-correlation matrix, and skewness \cite{papapanagiotou2017splendid}).

Finally, the sensor we use in this work includes two audio channels, left and
right. Based on visual inspection of the signals we have identified that the
left and right channels are practically equivalent, and can be used
interchangeably. However, we also perform some aggregation tests to confirm our
inspection conclusions. We examine the following cases:
\begin{enumerate}
\item Early fusion of features: for each window, the full feature set is
  extracted from each channel and the two feature vectors are then concatenated
  into a single feature vector
\item Late fusion of SVM scores: we train one SVM model per channel, and obtain
  two SVM scores per window (one from the left and one from the right channel
  classifier); we then aggregate them using (a) the $\max$ operator, or (b) by
  training a third SVM on the 2D ``feature vector'' of scores
\item We only use one channel, either the left or the right one
\end{enumerate}

It is important to note that the algorithm of \cite{papapanagiotou2017splendid}
is originally applied on audio signals with $2$ kHz sampling rate. In this work,
we adjust the thresholds, filters, and feature extraction parameters
accordingly, to accommodate the higher sampling rate of the smart-glasses.

\section{Evaluation framework and results}
\label{sec:experimental}

\subsection{Classifier training}

Our dataset includes $5$ subjects. To train the binary SVM classifier we perform
leave-one-subject-out experiments. We use the radial-basis function (RBF)
kernel. For each training, we select the hyperparameters $C$ of the SVM
classifier and $\gamma$ of the RBF kernel by performing cross-validation on the
training data. The hyperparameters space is traversed using Bayesian
optimization (instead of a plain grid-search). Finally, to reduce the
computational time we do not use all the available training data each time, but
we randomly sample only $500$ positive and $500$ negative windows. Experimental
results with greater samples ($1,000$ and $2,000$ per class) yield similar
results but significantly increase the computational time.

\subsection{Evaluation framework}

Applying the algorithm on an audio recording yields a sequence of SVM
scores. Thresholding these scores (typically at $0$) yields binary labels that
correspond to chewing vs. non-chewing.

We first evaluate the effectiveness of the trained classifiers directly by
computing a binary confusion matrix based on the window labels. We compute
precision, recall, and F1-score for each subject and also the mean across
subjects.

We also extract eating events using the aggregation from chews to chewing bouts,
and then to eating events, as described in \cite{papapanagiotou2017}. The
aggregation is applied both on the ground truth and the detected chews. This
yields a sequence of ground truth eating sessions (i.e., start and stop
timestamps) and a sequence of detected eating sessions. We evaluate by
partitioning each recording duration into true positive (TP), true negative
(TN), false positive (FP), and false negative (FN) time intervals. We then
compute the same metrics as earlier. For this, we count the time that is marked
as eating from both the ground truth and detector as TP positive time, the time
that is marked as eating from ground truth and non-eating from the detector as
FN, and similarly for FP and TN time.

We also construct precision vs. recall plots by varying the decision threshold
of the SVM scores (after the filtering).

\subsection{Results and discussion}

We first compare the two channels using the early and late fusion methods
described in section \ref{sec:algorithm}. Early fusion yields an average (across
subjects) F1-score (on window based evaluation) of $0.657 \pm 0.078$. Late
fusion yields $0.647 \pm 0.063$ using the $\max$ operator and $0.64 \pm 0.076$
using a third SVM. Finally, using only the left channel we obtain an F1-score of
$0.657 \pm 0.079$ and using only the right channel $0.642 \pm 0.08$. Based on
these results, we may pick any of the two channels. All following analysis is
based on the right channel.

\begin{figure}
  \centering
  \includegraphics{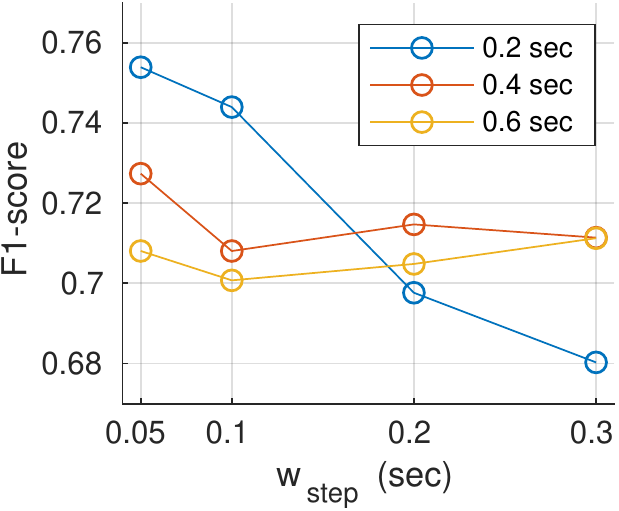}
  \caption[scale=.8]{Evaluation results for different window parameters (size
    and step): F1-score for window-based classification, for three different
    window sizes (the three colored lines) across different window steps.}
  \label{fig:window-study}
\end{figure}

We then examine the effect of the window size and step. The work of
\cite{papapanagiotou2017} uses two different window sizes: $0.2$ sec for some of
the features and $0.1$ sec for the rest. In this we follow the simpler approach
and use the same window size for all features; we test values $0.2$, $0.4$, and
$0.6$. We also test the following window step values: $0.05$, $0.1$, $0.2$,
$0.3$. Figure \ref{fig:window-study} presents the evaluation results in terms of
the F1-score for the window-level classification. The effectiveness benefits
from smaller window sizes. The value for $0.2$ sec window size and $0.3$ window
step is expected to be worse since the window step is larger than its size,
creating ``gaps'' of unused signal between successive windows. Based on these
results, we select the values of $0.2$ and $0.05$ for size and step.

\begin{figure}
  \centering
  \includegraphics[scale=1]{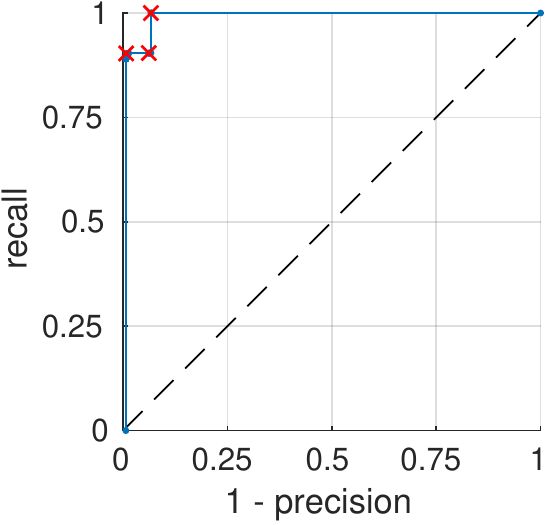}
  \caption{Precision-recall curve for duration-based evaluation of eating
    time. Each point is the average (across subjects) for a different threshold
    of SVM scores (default is $0$). Exact values for the three red points are
    shown in Table \ref{tab:results}. Area under curve (AUC) is $0.9874$.}
  \label{fig:roc}
\end{figure}

Finally, we formulate chews as ``pulses'' of the binary SVM score, and aggregate
to chewing bouts and then eating events and evaluate based on duration. We then
plot the precision-recall curve by varying the threshold for the SVM score
(default is $0$), to examine the limits of our approach in terms of precision
and recall. Figure \ref{fig:roc} shows the result. Results are very encouraging,
as the area under curve (AUC) is $0.9874$. We have chosen three points, one with
very high precision, one with very high recall, and one balanced, and show the
exact values in Table \ref{tab:results}. F1-score is equal or greater than
$0.92$.

\begin{table}
  \centering
  \caption{Evaluation results for selected points of the precision-recall curve
    (marked red in Figure \ref{fig:roc}).}
  \label{tab:results}
  \begin{tabular}{ccc}
    \toprule
    \textbf{precision} & \textbf{recall} & \textbf{F1-score} \\
    \midrule
    $0.9331$ & $1.0000$ & $0.9654$ \\
    $0.9382$ & $0.9041$ & $0.9208$ \\
    $0.9925$ & $0.9032$ & $0.9457$ \\
    \bottomrule
  \end{tabular}
\end{table}

\section{Conclusions}
\label{sec:conclusions}

In this work we examine the feasibility of using off-the-shelf,
commercially-available hardware for automatic eating detection. We perform
chewing detection based on audio captured by Bluetooth-enabled smart-glasses,
the Razer Anzu. We combine this choice with a robust, chewing-detection
algorithm from literature, and evaluate on an in-house yet challenging dataset
of over $6$ hours. Detection effectiveness for eating vs. non-eating time is
very promising, yielding F1-score above $0.92$ and as high as $0.965$ for the
best case. Future work includes making a variant of our dataset public to enable
direct comparison of similar approaches on this hardware, better leveraging the
two microphones and examine the possibility of identifying the side of chewing
(left, right, or middle), and finally evaluating on larger and more challenging
datasets.

\begin{acks}
  The work leading to these results has received funding from the European
  Community's Health, demographic change and well-being Program under Grant
  Agreement No. 965231, 01/04/2021 - 31/03/2025
  \url{https://rebeccaproject.eu/}.
\end{acks}

\printbibliography

\end{document}